\documentclass[runningheads,fleqn]{svmult}
\usepackage{makeidx}   % allows index generation
\usepackage{graphicx}  % standard LaTeX graphics tool
\usepackage{subeqnar}  % subnumbers individual equations
\usepackage{multicol}  % used for the two-column index
\usepackage{taphys}    % flushleft layout of captions etc.
\makeindex             % used for the subject index
\newcommand{\be}{\begin{equation}}      
\newcommand{\ee}{\end{equation}}      
      
\newcommand{\bef}{\begin{figure}}      
\newcommand{\eef}{\end{figure}}     
\def\spose#1{\hbox to 0pt{#1\hss}}      
\def\ltapprox{\mathrel{\spose{\lower 3pt\hbox{$\mathchar"218$}}      
 \raise 2.0pt\hbox{$\mathchar"13C$}}}      
\def\gtapprox{\mathrel{\spose{\lower 3pt\hbox{$\mathchar"218$}}      
 \raise 2.0pt\hbox{$\mathchar"13E$}}}      
\def\inapprox{\mathrel{\spose{\lower 3pt\hbox{$\mathchar"218$}}      
 \raise 2.0pt\hbox{$\mathchar"232$}}}

\begin{document}
\title*{Statistical physics for complex cosmic structures}
\toctitle{Statistical physics for complex cosmic structures}
% allows explicit linebreak for the table of content
%
%
\titlerunning{Statistical physics for complex cosmic structures}
% allows abbreviation of title, if the full title is too long
% to fit in the running head
%
\author{Luciano Pietronero \inst{1}
\and Francesco Sylos Labini \inst{2}}
\authorrunning{Pietronero \& Sylos Labini}
% if there are more than two authors,
% please abbreviate author list for running head
%
%
\institute{Physics Department University of Rome ``La Sapienza'',
P.le A. Moro 2, 00185 Rome Italy
\and 
Universit\'{e} de Paris-Sud,
     Laboratoire de Physique Theorique 
     B\^{a}timent 210,\\
     F-91405 Orsay Cedex, France}

\maketitle              % typesets the title of the contribution

\begin{abstract}
Cosmic structures at large scales represent the earliest and most
extended form of matter condensation. In this lecture we review the
application of the methods and concepts of modern statistical physics
to these structures. This leads to a new perspective in the field
which can be tested by the many new data which are appearing in the
near future. In particular, galaxy structures show fractal correlation
up to the present observational limits. The cosmic microwave
background radiation, which should trace the initial conditions from
which these structures have emerged through gravitational dynamics, is
instead extremely smooth. Understanding the relation between the
complex galaxy structures and the smooth microwave background
radiation represents an extremely challenging problem in the field of
structure formation.
\end{abstract}

\section{Introduction}

The specific issue which initiated our activity at the interface
between statistical physics and cosmology, was the study of the
clustering properties of galaxies as revealed by large redshift
surveys, a context in which concepts of modern statistical physics
(e.g.  scale-invariance, fractality..)  found a ready application
\cite{cp92,slmp98,cfa2}. In recent years we have broadened considerably the
range of problems in cosmology which we have addressed, treating in
particular more theoretical issues about the statistical properties of
standard primordial cosmological models \cite{book,lebo,glass} and the
problems of formation of non-linear structures in gravitational N-body
simulations \cite{nbody1,nbody2}.  What is common to all this activity
is that it is informed by a perspective and methodology which is that
of statistical physics: in fact we believe that this represents an
exciting playground for statistical physics, and one which can bring
new and useful insights into cosmology.  In this lecture we briefly
review the main points and we refer the interested reader to the
recent monograph on the subject \cite{book}.

\section{Galaxy structures} 

The question which first stimulated our interest in
cosmology has been the correlation properties of the observed
distribution of galaxies and galaxy clusters \cite{cp92,slmp98,cfa2}.  
It is here that the
perspective of a statistical physicist, exposed to the developments of
the last decades in the description of intrinsically irregular
structures, is radically different from that of a cosmologist for whom
the study of fluctuations means the study of small fluctuations {\it
about a positive mean density}. And it is here therefore that the
instruments used to describe strong irregularity, even if limited to a
finite range of scales, offer a wider framework in which to approach
the problem of how to characterize the correlations in galaxy
distributions.  Analyzing galaxy distributions in this wider context
means treating these distributions without the {\em a priori}
assumption of homogeneity, i.e. without the assumption that the finite
sample considered gives, to a sufficiently good approximation, the
true (non-zero) mean density of the underlying distribution of
galaxies. While this is a simple and evident step for a statistical
physicist, it can seem to be a radical one for a cosmologist 
\cite{rees,pee93}. After
all the whole theoretical framework of cosmology (i.e. the
Friedmann-Robertson-Walker --- FRW --- solutions of general relativity) is
built on the assumption of an homogeneous and isotropic distribution
of matter. Our approach is an empirical one, which surely is
appropriate when faced with the characterization of data. Further it
is evidently important for the formulation of theoretical explanations
to understand and characterize the data correctly.

\begin{figure}
\centerline{
\includegraphics*[width=.75\textwidth]{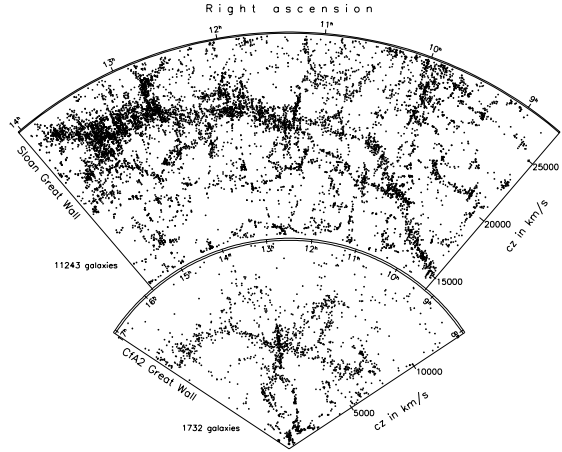}}
\caption[]{Sloan Great Wall compared to CfA2 Great Wall 
at the same scale..
Redshift distances $cz$ are indicated (distances are 
simply given by $cz/100$ Mpc/h). The Sloan slice is $4$ degrees
wide, the CfA2 slice is $12$ degrees wide to make both slices
approximately the same physical width at the two walls. 
(From \cite{logmap} - see also 
{\tt http://www.astro.princeton.edu/$\sim$mjuric/universe/)}}
\label{sdss}
\end{figure}

The maps probing the distribution of galaxies revealed in the first
large three dimensional surveys, which were published in the eighties,
structures - super-clusters, walls, voids, filaments - at scales much
larger than had been suspected from the previous (fairly isotropic)
angular data (see Fig.\ref{sdss}). 
The simple visual impression of such three-dimensional data -
apparently showing large fluctuations up to the sample sizes - gives a
strong {\it prime face} case for an analysis which does not assume the
underlying distribution is uniform (at the scales probed), but rather
encompasses the possibility that it may be intrinsically
irregular. The quantitative results obtained with the standard
analysis performed on these samples, which works with statistics which
build in the assumption of homogeneity, give further evidence in this
direction.

In simple terms, the standard statistical framework used in cosmology
is the one developed for the studies of regular distributions as
fluids: one assumes that the average density $\langle n \rangle$ is
well defined within a given sample and thus one focuses on the
description of fluctuations around such a quantity (where $\langle
... \rangle$ is the unconditional ensemble average). In order to use a
normalized (or reduced) two-point correlation function one defines
\be
\label{xi}
\xi(r) = \frac{\langle n(r) n(0) \rangle} {\langle n \rangle^2} -1
\ee
so that $\xi(r) \ne 0$ implies the presence of correlations,
while $\xi(r)=0$ is for a purely Poisson (uncorrelated) distribution:
when $\xi(r) \gg 1$ the regime is of strong-clustering
while $\xi(r) < 0$ describes anti-correlations.  
However such a definition does not include the possibility that 
the average density $\langle n \rangle$ could  not be a well-defined
quantity, i.e. that it could  
be a function of the sample size. This can considered by using
the conditional average density defined as 
\be
\label{gamma}
\langle n(r) \rangle_p = 
\frac{\langle n(r) n(0) \rangle} {\langle n \rangle}
\ee
which gives the average density around an occupied point of the
distribution
\cite{cp92,slmp98,book} (where $\langle ... \rangle_p$ is the 
conditional  ensemble average). Clearly there is a simple relation which 
links these two functions 
\[
\langle n(r) \rangle_p = \langle n \rangle  (1+\xi(r)) \;,
\]
which is physically meaningful only in the case when the average
density is a well defined quantity. Together with this simple change
of statistical description of correlations one should consider however
a subtle but important point related to the non-analytical character
of fractal objects.

Fractal geometry, has allowed us to classify and study a large variety
of structures in nature which are intrinsically irregular and
self-similar \cite{man82}.  The {\it metric} dimension is the most
important concept introduced to describe these intrinsically irregular
systems.  Basically it measures the rate of increase of the ``mass''
of the set with the size of the volume in which it is measured: In
terms of average density around an occupied point (averaged over all
occupied points) if it scales as
\be
\label{gammafra}
\langle n(r) \rangle_p \sim r^{D-3} 
\ee
then the exponent $D$ is called the {\it mass-length dimension} and it
is $0 < D \le 3$ (in the three dimensional Euclidean space, where
$D=3$ corresponds to the uniform case). In this situation the reduced
correlation function (Eq.\ref{xi}) looses its physical meaning
although its estimator can be defined in a finite sample. This is so,
because its amplitude depends on the sample size as the average
density does. In fact one may show that in this case the estimator of
$\xi(r)$ in a finite spherical sample of size $R_s$ becomes (in an infinite
volume it cannot be defined as a fractal is asymptotically empty,
i.e. $\langle n \rangle =0$)
\be 
\label{xi_f} 
\xi_E(r) = \frac{D}{3} \left( \frac{r}{R_s}\right)^{D-3} -1 \;.
\ee
The use of the conditional average density we consider
(Eqs.\ref{gamma}-\ref{gammafra}) encompasses both some irregular
distributions (simple fractals) and uniform distributions with small
scale clusterization, in particular those described by standard
cosmological models.  If the sample size is much larger than the
homogeneity scale $\lambda_0$ one can detect using $\langle n(r)
\rangle_p$ the existence and location of $\lambda_0$.  If, on the
other hand the sample's size is smaller than $\lambda_0$ the
conditional density allows the determination of the fractal exponent
characterizing the clustering on these scales.

While the reduced correlation function $\xi(r)$ has been found to show
consistently a simple power-law behavior characterized by the same
exponent in the regime of strong clustering, there is very
considerable variation between samples, with different depth and
luminosity cuts, in the measured {\it amplitude} of
mean-density-normalized correlation function \cite{sdss2}.  This
variation in amplitude is usually ascribed {\it a posteriori} to an
intrinsic difference in the correlation properties of galaxies of
different luminosity (``luminosity bias'', or ``luminosity
segregation''). It may, however, have a much simpler explanation in
the context of irregular distributions. As mentioned above, in a
simple fractal (at least within the sample scale), for example, the
density in a finite sample decreases on average as a function of
sample size; samples of increasingly bright galaxies are in fact
generically of greater mean depth, which corresponds to an increasing
amplitude of the correlation function $\xi(r)$ normalized to the
``apparent'' average density in each sample (see Eq.\ref{xi_f}).

\begin{figure}
\centerline{
\includegraphics*[width=.55\textwidth]{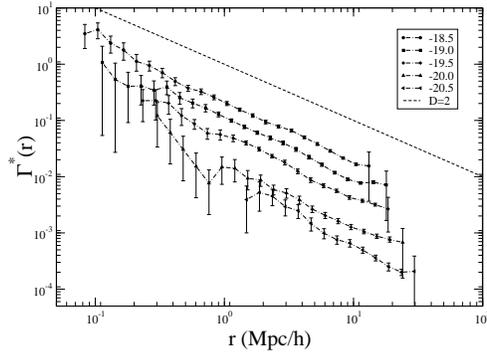}}
\caption[]{The  
estimator of the conditional average density $\Gamma^*(r)$ in the
different samples of the CfA2-South galaxy catalog. (Adapted from
\cite{cfa2}).}
\label{gammacfa2}
\end{figure}

With this motivation we have applied these simple statistical methods which
allow a characterization of galaxy clustering, irrespective of whether
the underlying galaxy distribution is homogeneous or irregular at the
sample size.  Up to scales of tens of Mpc, we have found that galaxy
distributions in many different surveys \cite{slmp98,cfa2} show
fractal properties corresponding to $D \approx 2$ with no robust
evidence for homogeneity (see Fig.\ref{gammacfa2}).
It is on this important issue that future surveys (e.g. the Sloan
Digital Sky Survey --- SDSS --- \cite{sdss}) will allow us to place
much tighter constraints.  We stress however the following important
points: if the homogeneity scale is determined to be finite, the
fractal-inspired analysis remains valid in two respects: (i) it is the
unique way to detect the homogeneity scale itself without a priori
assumptions, and (ii) it gives the right framework to obtain a full
geometrical and statistical description of the strongly clustered
region \cite{book}. This result has caused a passionate debate in the
field because it is in contrast with the usual assumption of large
scale homogeneity which is at the basis of most theories. Our basic
conclusion is that, given the fractal nature of galaxy clustering on
small scales, one should change the general perspective with respect
to the problem of non-linear structure formation. Note that this is
the case even it turns out that there is crossover to homogeneity
because, at least in a certain range of scales, that in which there is
strong clustering, structures have a fractal nature and this cannot be
described as perturbations to a smooth fluid.  It can be interesting
to note that some cosmologists (e.g. \cite{rees}) have confirmed our
measurements about fractality at scales of order tens Mpc, the debate
about the transition to homogeneity concerning scales of one order of
magnitude larger.  The discussion about the homogeneity scale is
indeed an hot topic in contemporary science, touching many deep and
important key-issues in different fields of physics, from large scale
astrophysics to gravitation, to the question of the nature and amount
of dark matter.  To these topics it is dedicated a recent book
\cite{yuripekka}:  Its main aim is in fact to address, in a novel and
open way, the problem of structures in cosmology, with a reach
historical perspective and stressing the open questions which arise
from the consideration of their intrinsic complexity.

One of the reason of the resistance to our ideas in cosmology, is that
we do not present neither an alternative theory or a model nor a
physical theory for the fractal growth phenomena in the gravitational
case: this is due to the fact that this issue is indeed extremely
difficult as any out-of-equilibrium irreversible dynamical growth
process. Actually the study of such processes is at the frontier of
the theoretical efforts in statistical physics. Let us, however, note
briefly two common {\it theoretical} objections to this approach. One
is that fractals are incompatible with what is called the
``cosmological principle'', by which it is meant that there is no
privileged point or direction in the universe. This is a simple
misconception about fractals. These irregular distributions are in
keeping with this principle in exactly the way as inhomogeneities
treated in the standard framework of perturbed FRW models, i.e. they
are {\it statistically} stationary and isotropic distributions in
space in the sense that their statistical properties are invariant
under translation and rotation in space. Another common objection is
that there is an inconsistency in using the Hubble law, which is used
to convert redshift to physical distance, if one does not assume
homogeneity. This objection forgets that the Hubble law is an
established empirical relation, independent of theories explaining
it. Further it can be noted that what we are concerned with is the
distribution of {\it visible} matter: given that standard cosmological
models describe a universe whose energy density is completely
dominated by several non-visible components, reconciling the two is
not in principle impossible
\cite{pwa00}.

\section{Primordial density fields} 

As already mentioned, the standard interpretation of galaxy and
cluster correlations results from the $\xi(r)$ analysis and can be
summarized as follows: (i) Power-law behavior in the regime of strong
clustering ($\xi \gg 1$) (ii) Amplitude which changes from sample to
sample. These observations of the varying amplitude are ascribed to
real physics, rather than being simply finite size effects as in the
alternative explanation we present. Generically this goes under the
name of ``bias'', which is then simply taken to mean any difference
between the correlation properties of any class of object (galaxy of
different luminosity or morphology, galaxy cluster, quasar...) and
that of the ``cold'' dark matter (CDM) which, in standard models,
dominates the gravitational clustering dynamics in the universe. The
case that such a difference manifests itself as an overall
normalization of $\xi(r)$, as in the observations we have been
discussing, is known as ``linear bias''.  In order to understand the
problematic aspects of this interpretation, one should consider an
important feature common to all standard theoretical models of
cosmological density fields, like CDM.

First let us briefly summarize the main properties of these fields.  Matter
distribution in the early universe is supposed to be well represented
by a continuous Gaussian field with appropriate correlation
properties. According to the standard picture of the Hot-Big-Bang
scenario \cite{pee93} matter density fluctuations had interact with
radiation at early times leaving therefore an imprint of its
correlation properties of the cosmic microwave background radiation
(CMBR) anisotropies, in particular determining an important property
at large scales. This is the so-called Harrison-Zeldovich (HZ)
condition on the power spectrum (PS) $P(k)$ (Fourier Transform of
$\xi(r)$) of these tiny fluctuations with respect to perfect
uniformity \cite{glass}: $P(k) \sim k$ at small $k$ (large scales).
We have noticed in \cite{glass} that this condition implies that their
primary characteristic is to show surface mass fluctuations, which are
the most depressed fluctuations possible for any stochastic
distribution.  In a simple classification of all stationary stochastic
processes into three categories, we highlighted with the name
``super-homogeneous'' the properties of the class to which models like
this belong: let us briefly discuss this point.

A simple description of these systems can be the following.  An
uncorrelated Poisson system (or with positive and finite range of
correlations) has at large scale $\xi(r) =0 $ and hence the PS is
$P(k) \sim const$: for this reason it satisfies the condition
\[
\int_0^{\infty} \xi(r) r^2 dr \sim P(0) = const.
\]
In this situation mass fluctuations in spheres grows as $\langle
\Delta M^2 (R) \rangle \sim R^3$ (Poisson fluctuations): this is, for
example a perfect gas at thermodynamical equilibrium.  On the other
hand, one may find systems with a power-law correlation function as in
thermodynamical critical phenomena. In this situation $\xi(r) \sim
r^{-\gamma}$ where $0 < \gamma < 3$, $P(k) \sim k^{-(3-\gamma)}$ and
\[
\int_0^{\infty} \xi(r) r^2 dr  \sim P(0) = \infty \;: 
\]
The correlation length diverges and the system 
has critical features is that mass fluctuations growth faster than in the 
Poisson case, i.e. 
$\langle \Delta M^2 (R) \rangle \sim R^{6 - \gamma}$.
Finally there are systems such that 
\be
\label{p0}
\int_0^{\infty} \xi(r) r^2 dr \sim P(0) = 0 
\ee
and mass fluctuations in spheres go  as $\langle
\Delta M^2 (R) \rangle \sim R^2$: 
In statistical physics language they are well described as glass-like
or long-range ordered configurations. For example the generation of
points distributions in three dimensions with properties similar to HZ
ones encountered in statistical physics is represented by the
one-component system of charged particles in a uniform background
\cite{lebo}.  This latter system, appropriately modified, can produce
equilibrium correlations of the kind assumed in the cosmological
context.  These systems are characterized by a long-range balance
between correlations and anti-correlations (Eq.\ref{p0}) and show the
most depressed possible fluctuations in any stochastic distribution
\cite{glass}.

The HZ type spectrum was first given a special importance in cosmology
with arguments for its ``naturalness'' as an initial condition for
fluctuations in the framework of the expanding universe
cosmology. Basically it satisfies the global condition on large scale
mass fluctuations $P(0)=0$, which is the only one compatible with the
FRW metric, as it avoids the divergence of the fluctuations of the
gravitational potential (as it happens in the simple Poisson case
\cite{glass}). It subsequently gained in importance with the advent of
inflationary models in the eighties, and the demonstration that such
models quite generically predict a spectrum of fluctuations of this
type. Since the early nineties, when the COBE experiment (and then
more recently WMAP
\cite{wmap}) measured for the first time fluctuations in the
temperature CMBR at large angular separations, and found results
consistent with the predictions of models with a HZ spectrum at such
scales, the HZ type spectra have become a central pillar of standard
models of structure formation in the universe.

Because of the highly irregular nature of structure at small scales,
standard models with super-homogeneous features cannot be used even at
zeroth order to describe these observed structures. This does not
mean, however, that these models cannot describe successfully galaxy
structures: but to establish whether they can, it must first be shown
from observations that there is a clear crossover toward homogeneity
i.e. a scale beyond which the average density becomes a well-defined
(i.e. sample-independent) positive quantity. These models then predict
that, on much larger scales (e.g. $> 100$ Mpc), galaxy structures
should present the super-homogeneous character of the HZ type
PS. Indeed this should in principle be a critical test of the paradigm
linking the measurements of CMBR on large scales to the distribution
of matter. Observationally a crucial question is the feasibility of
measuring the transition between these regimes directly in galaxy
distributions. In this context one has to consider an important
element: the galaxy distribution is a discrete set of objects whose
properties are related in a non-trivial way to the ones of the
underlying continuous field. To understand the relation between the
two, one has to consider the additional effects related to sampling
the continuous field. This is intimately related to the problem of
``biasing'' between the distribution of visible and dark matter which,
as above mentioned, is usually invoked to explain a posteriori the
observational features of $\xi(r)$.

\begin{figure}
\centerline{
\includegraphics*[width=.75\textwidth]{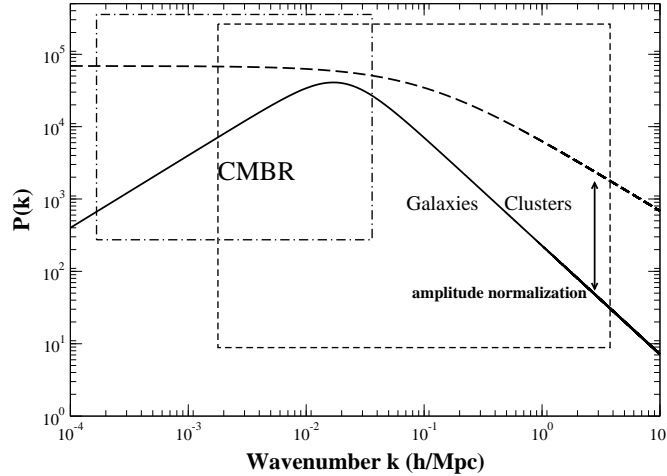}
}
\caption[]{In this figure we show a typical CDM type power spectrum
(solid line) and the range associated to the primary observational
constraints. The left hand side, the Harrison Zeldovich part of the PS
($P(k) \sim k$), is constrained by observations of the anisotropies of
the CMBR (dash-dotted box). Current galaxy and galaxy cluster surveys
gives constraints at smaller scales (dashed box). The normalization of
the amplitude of the galaxy or cluster PS to the one observed in the
CMBR is fundamentally important. It is usually determined by a {\it
linear} re-scaling on the y-axis, ascribed to the effect of bias. This
simple assumption is not consistent with the canonical model for the
biasing of a Gaussian field, which introduces a non-linear distortion
both at small and large wave-number. This is illustrated by the dashed
line, which shows what is actually obtained for the PS of the biased
field. On small scales (large $k$) there is a non-linear distortion
and at large scales (small $k$) the behavior is typical of a
substantially Poisson system with $P(k) \sim const$. 
(From \cite{book}). }
\label{figbias}
\end{figure}

Sampling a super-homogeneous fluctuation field may change the nature
of correlations \cite{bias}. The reason can be found in the property
of super-homogeneity of such distribution: the sampling, as for
instance in the so-called ``bias model'' (selection of highest peaks
of the fluctuations field) necessarily destroys the surface nature of
the fluctuations, as it introduces a volume (Poisson-like) term in the
variance. The ``primordial'' form of the power spectrum is thus not
apparent in that which one would expect to measure from objects
selected in this way. This conclusion should hold for any generic
model of bias \cite{bias}. If a linear amplification is obtained in
some regime of scales (as it can be in certain phenomenological models
of bias) it is necessarily a result of a fine-tuning of the model
parameters. The study of different samplings and of the correlation
features invariant under sampling represents an important issue in
relation to the comparison of observations of galaxy structures, or
distributions given by N-body simulations with primordial fluctuations
(CMBR anisotropies) and theoretical models (see Fig.\ref{figbias}).

\section{The problem of gravitational structure formation}

\begin{figure}
\centerline{
\includegraphics*[width=.75\textwidth]{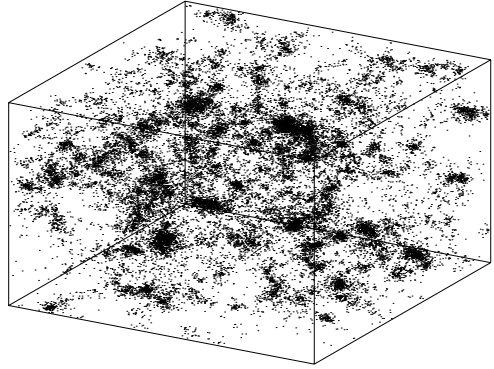}
}
\caption[]{This is a distribution obtained by starting 
from a Poisson particle configuration with zero initial 
velocity dispersion and by evolving it with gravity and by
considering periodic boundary conditions. One may see that 
non-linear strong clustering is formed  and its explanation
is  the central task of theoretical modeling. 
(Courtesy of T. Baertschiger).
}
\label{poisson}
\end{figure}

In cosmology the main instrument for treating the {\it theoretical}
problem of gravitational clustering is numerical (in the form of
N-body simulations ---NBS---), and the analytic understanding of this
crucial problem is very limited.  Other than in the regime of very
small fluctuations where a linear analysis can be performed, the
available models of clustering are essentially phenomenological models
with numerous parameters which are fixed by numerical simulation. An
NBS has these main features: (i) given $N$ particles in a
volume $V$, each particle moves according to the force given by the sum
of the other $N-1$ particles and of all their replicas (due to the use
of periodic boundary conditions). (ii) The equation of motions are
integrated by using a leap-frog algorithm. (iii) Suitable techniques
allow an appropriate summations (which makes the code faster) of far
away particles contributions to the gravitational force. (iv) One may,
or may not, consider the presence of space expansion. A typical result
of the non-linear evolution is shown in Fig.\ref{poisson}.

A central issue in the context of cosmological NBS is to relate the
formation of non linear structures to the specific choice of initial
conditions used: this is done in order to constrain models with the
observations of CMBR anisotropies, which are related to the initial
conditions, and of galaxy structures, which give instead the final
configuration of clustered matter. Standard primordial cosmological
theoretical density fields, like the CDM case, are Gaussian and made
of a huge number of very small mass particles, which are usually
treated theoretically as a self-gravitating collision-less fluid: this
means that the fluid must be dissipation-less and that two-body
scattering should be negligible.  The problem then being in which
limit NBS, based on particle dynamics, are able to reproduce the two
above conditions \cite{nbody1,nbody3}. In this context one has to
consider the issue of the physical role of particle fluctuations in
the dynamics of NBS. In fact, in the discretization of a continuous
density field one faces two important limitations corresponding to the
new length scales which are introduced. One the one hand a relatively
small number of particles are used. This introduces a mass scale which
is the mass of these particles. In typical cosmological NBS, this mass
is of the order of a galaxy and hence many orders of magnitude larger
than the microscopic mass of a CDM particle. Furthermore, it
introduces a new length scale given by the average distance between
nearest neighbor particles. On the other hand one must regularize the
gravitational force at small scales in order to avoid numerical
problems and typical small scale effects due to the discrete nature of
the particles: given that the smoothening length is typically smaller
than the initial inter-particle separation, it is not evident that one
is effectively reproducing a dynamics where particles play the role of
collision-less fluid elements.  It is in this sense that one talks
about the role of discreteness in NBS: that strong scattering between
nearby particles are produced by the discretization and they should be
considered artificial and spurious with respect to the dynamical
evolution of a self-gravitating fluid. This point has been considered
by many authors and they all show that discreteness has some influence
on the formation of the structures \cite{bottaccio,nbody3}. Indeed,
discreteness may play an important role in the early times formation
of non-linear structures.  How discrete effects are then ``exported''
toward large scales, if they are at all, is then an important but
difficult problem to be understood. In other words the problem is that
of understanding whether large non-linear structures, which at late
times contain many particles, are produced solely by collision-less
fluid dynamics, or whether the particle collisional processes are
important also in the long-term, or whether they are made by a mix of
these two effects \cite{nbody3}.

For example in \cite{nbody2} we have presented an analysis of
different sets of gravitational NBS all describing the dynamics of
discrete particles with a small initial velocity dispersion. They
encompass very different initial particle configurations, different
numerical algorithms for the computation of the force, with or without
the space expansion of cosmological models. Despite these differences
we find in all cases that the non-linear clustering which results is
essentially the same, with a well-defined simple power-law behavior in
the conditional density in the range from a few times the lower
cut-off in the gravitational force to the scale at which fluctuations
are of order one. We have argued, presenting quantitative evidence,
that this apparently universal behavior can be understood by the
domination of the small scale contribution to the gravitational force,
coming initially from nearest neighbor particles \cite{nbody3}. A more
quantitative description of this dynamics is evidently needed, with
the principal goal of understanding the specific value observed of the
exponent. In the cosmological literature (see e.g. \cite{pee93}) the
idea is widely dispersed that the exponents in non-linear clustering
are related to that of the initial PS of the small fluctuations in the
CDM fluid, and even that the non-linear two-point correlation can be
considered an analytic function of the initial two-point correlations.
The models used to explain the behavior in the non-linear regime
usually involve both the expansion of the Universe, and a description
of the clustering in terms of the evolution of a continuous fluid. We
have argued that the exponent is universal in a very wide sense, being
common to the non-linear clustering observed in the non-expanding
case. It would appear that the framework for understanding the
non-linear clustering must be one in which discreteness (and hence
intrinsically non-analytical behavior of the density field) is
central, and that the simple context of non-expanding models should be
sufficient to elucidate the essential physics.

There are then some basic questions which remain unanswered: does
gravitational dynamics give rise to fractal clustering ?  And, in case, from
what initial conditions ? We believe there is much unexplored space
for a statistical physics approach to the problem of gravitational
clustering, which should be able to shed light on the more general
characteristics of non-linear structure formation. 
\bigskip

%\section*{Acknowledgments}
We warmly thanks our collaborators and particularly T. Baertschiger,
A. Gabrielli and M. Joyce. FSL acknowledges the support of a Marie
Curie Fellowship HPMF-CT-2001-01443.

%%%%%%%%%%%%%%%%%%%%%%%%%%%%%%%%%%%%%%%%%%%%%%%%%%%%%%%%%%%%%%%%%%%%%%%

%INDEX%%%%%%%%%%%%%%%%%%%%%%%%%%%%%%%%%%%%%%%%%%%%%%%%%%%%%%%%%%%%%%%
% Please code your entries to include a "mutual" subject index in the
% standard syntax. For your own purposes you may print your
% "personal" index by using the following commands:
%
%\clearpage
%\addcontentsline{toc}{section}{Index}
%\flushbottom
%\printindex
%%%%%%%%%%%%%%%%%%%%%%%%%%%%%%%%%%%%%%%%%%%%%%%%%%%%%%%%%%%%%%%%%%%%%

\end{document}